\documentclass[aps,pre,preprint,superscriptaddress]{revtex4-1}

\DeclareUnicodeCharacter{2212}{-}
\usepackage{graphicx,amssymb, verbatim,amsmath,siunitx, color,float,soul,makecell,xcolor}
\usepackage{natbib,url,enumerate}
\usepackage{tabularx}
\usepackage{bigints}
\usepackage[colorlinks,allcolors = blue,bookmarksopen,bookmarksnumbered]{hyperref}

\usepackage{hyperref}
\usepackage{silence}
\definecolor{colsq}{rgb}{0,0.4470,0.7410}
\definecolor{coltr}{rgb}{0.8500,0.3250,0.0980}
\definecolor{coldm}{rgb}{0,0.4470,0.7410}
\definecolor{colbtr}{rgb}{0.9290,0.6940,0.1250}

\newcommand{\be}{\begin{equation}}
\newcommand{\ee}{\end{equation}}
\newcommand{\bse}{\begin{subequations}}
\newcommand{\ese}{\end{subequations}}
\newcommand{\bea}{\begin{eqnarray}}
\newcommand{\eea}{\end{eqnarray}}
\newcommand{\ba}{\begin{array}}
\newcommand{\ea}{\end{array}}

\begin{document}

\title{Spectral Energy Transfers in Domain Growth Problems}

\author{Pradeep Kumar Yadav}
\email{pradeep18.duphysics@gmail.com}
\affiliation{School of Physical Sciences, Jawaharlal Nehru University, New Delhi--110067, India.}
\author{Mahendra Kumar Verma}
\email{mkv@iitk.ac.in}
\affiliation{Department of Physics, Indian Institute of Technology Kanpur, Kanpur--208016, India.}
\author{Sanjay Puri}
\email{purijnu@gmail.com}
\affiliation{School of Physical Sciences, Jawaharlal Nehru University, New Delhi--110067, India.}

\begin{abstract}
In the domain growth process, small structures gradually vanish, leaving behind larger ones. We investigate spectral energy transfers in two standard models for domain growth: (a) the {\it Cahn-Hilliard} (CH) equation with conserved dynamics, and (b) the {\it time-dependent Ginzburg-Landau} (TDGL) equation with non-conserved dynamics. The nonlinear terms in these equations dissipate fluctuations and facilitate energy transfers among Fourier modes. In the TDGL equation, only the $\phi(\mathbf{k} = 0, t)$ mode survives, and the order parameter $\phi(\mathbf{r},t)$ approaches a uniform state with $\phi = +1$ or $-1$. On the other hand, there is no dynamics of the $\phi(\mathbf{k} = 0, t)$ mode in the CH equation due to the conservation law, highlighting the different dynamics of these equations.
\end{abstract}

\maketitle
 
\section{Introduction}
\label{s1}

There has been intense interest in the process of \textit{domain growth} or \textit{coarsening}, which occurs when a two-phase mixture, initially in a homogeneous phase, is quenched below the critical coexistence temperature~\cite{bray2002theory,pw09,dp04}. The quench renders the disordered state thermodynamically unstable. The subsequent evolution is characterized by the emergence and growth of domains, which are enriched in one of the constituents of the mixture. In spite of significant progress in this field, the  relaxation to the asymptotic state still needs to be better understood. We have recently introduced \cite{verma2023nonlinear} spectral energy transfers, which are routinely used in studies of turbulence, as a novel technique to gain insights on the coarsening process. We illustrated this technique by applying it to the 1-dimensional $(D=1)$ \textit{time-dependent Ginzburg-Landau} (TDGL) equation or \textit{Model A}~\cite{hohenberg1977theory}. The present paper is a follow-up of Verma et al. \cite{verma2023nonlinear}. Here, we will apply the energy transfer technique to the \textit{Cahn-Hilliard} (CH) equation or \textit{Model B}~\cite{hohenberg1977theory}. The CH equation describes the case with conserved kinetics (e.g., phase separation dynamics of an AB mixture), whereas the TDGL equation corresponds to nonconserved kinetics (e.g., ordering of a ferromagnet). In the theory of domain growth processes, there is a very good (though approximate) understanding of the nonequilibrium evolution with nonconserved order parameter, as described by the domain growth law, scaling behavior of the domain morphology, etc.~\cite{bray2002theory,pw09,dp04}. Unfortunately, our understanding of the conserved case is far poorer, in spite of huge efforts in this direction. Perhaps the most important work in this context still remains the classic \textit{Lifshitz-Slyozov} (LS) paper \cite{lifshitz1961kinetics} written 60 years ago! The LS paper addresses the limit where one of the components is present in a vanishingly small fraction. However, the experimentally interesting limit is one where the two components are present in comparable fractions. This results in a bicontinuous morphology of the phase-separating system. In this paper, we introduce and use energy transfer methods to study evolution dynamics in the CH equation in $D=1,2$. For completeness, we also present analogous results for the $D=2$ TDGL equation, which was not addressed in our previous paper \cite{verma2023nonlinear}.

A common theme between turbulence and coarsening is the multi-scale interactions that lead to structure formation. There have been some recent works which have applied concepts from turbulence (e.g., energy transfers) to study other examples of pattern dynamics. Bratanov et al. \cite{bratanov2015new} investigated the turbulent motion of active fluids driven by convective nonlinearity of the Navier–Stokes type and  coarsening nonlinearity. These authors studied the energy transfers and dissipation between different scales and power law spectra. In their model, existence of a second nonlinearity provides additional freedom to the system to self-organize into states without the need for external fine-tuning. Recently, Mukerjee et al. \cite{Ray23} revisited the problem of turbulence in active fluids, and analyzed the variable energy fluxes that lead to deviations from Kolmogorov's $k^{-5/3}$ energy spectrum. Further, Perlekar et al. \cite{Perlekar2014spinodal,perlekar2017arrested} investigated coarsening arrest in binary fluid mixtures, including an inverse cascade of energy in \textit{Model H}. Rana, Perlekar and others~\cite{rana2022phasetonertu,rana2020coarseningTonertu} scrutinized the spectral properties of $D=2,3$ incompressible Toner-Tu equations. 

This paper is organized as follows. In Sec.~\ref{s2}, we introduce the CH and TDGL equations, and the formalism for nonlinear energy transfer and dissipation. In Sec.~\ref{s3}, we provide simulation details. In Sec.~\ref{s4}, we present our main simulation results for the CH equation in $D=1,2$. In Sec.~\ref{s5}, we briefly discuss numerical results for the TDGL equation in $D=1,2$. In Sec.~\ref{s6}, we conclude this paper with a summary and discussion.

\section{Energy Transfers and Dissipation in Coarsening}
\label{s2}

In this section, we will discuss energy transfers in the CH and TDGL equations.
 
\subsection{Cahn-Hilliard Equation}
\label{Methodology_II_cahn_hilliard}

Let us first consider the CH equation, which describes diffusion-driven phase-separation kinetics in a binary (AB) mixture. The volume fractions of A and B at time $t$ are $n_{A}({\bf r},t)$ and $n_{B}({\bf r},t)$, respectively. The order parameter is defined as  $\phi({\bf r},t) = n_{A}({\bf r},t) - n_{B}({\bf r},t)$. Initially, the system is prepared in a disordered state at a high temperature, where $\phi({\bf r},0) = \phi_{0} + \delta \phi ({\bf r},0)$. Here, $\delta \phi ({\bf r},0)$ denotes randomly-distributed small-amplitude fluctuations. At $t=0$, the system is rapidly quenched below its critical temperature and evolves towards its new equilibrium phase. This system is subject to the constraint of mass conservation $V^{-1} \int d{\bf r}~\phi({\bf r},t)= \phi_0$, where $V$ is the volume. Thus, $\phi({\bf r},t)$ obeys the continuity equation: 
\begin{eqnarray}
\frac{\partial \phi({\bf r},t)}{\partial t}  &=& -{\bf \nabla} \cdot \bf{J}({\mathbf r}, t) \nonumber \\
&=& {\bf \nabla} \cdot \left[ M {\bf \nabla} \left( \frac{\delta {\mathcal F[\phi]}}{\delta \phi} \right) \right]. 
\label{continuity}
\end{eqnarray}
Equation~(\ref{continuity}) is known as the CH equation, with $M$ being the mobility. In dimensionless units, the form of the Ginzburg-Landau (GL) free energy functional $\mathcal{F}$ is~\cite{bray2002theory,pw09} 
\begin{eqnarray}
\mathcal{F}[\phi] = \int d{\bf{r}} \left( -\frac{\phi^{2}}{2} + \frac{\phi^{4}}{4}+\frac{\left|{\nabla}\phi\right|^{2}}{2} \right) .
\label{functional_energy}
\end{eqnarray}
 The first two terms on the right-hand-side (RHS) of  Eq.~(\ref{functional_energy}) represent the bulk free energy density, and the square-gradient term models the penalty due to the spatial variation of $\phi({\bf r},t)$. The CH dynamics drives the order parameter to its asymptotic values $\phi = \pm 1$, corresponding to  A-rich and B-rich phases, respectively. The evolution of $\phi({\bf r},t)$ from Eqs.~(\ref{continuity})-(\ref{functional_energy}) is  
\begin{eqnarray}
\frac{{\partial \phi({\bf r},t)}}{\partial t} &=& -\nabla^2 \left(\phi-\phi^3 +  \nabla^2\phi\right).
\label{chcspace}
\end{eqnarray}

For the analysis of energy flows between different spatial scales by various terms in Eq.~(\ref{chcspace}), we write it in Fourier space:  
\begin{eqnarray}
 \frac{\partial\hat{\phi}({\bf k},t)}{\partial t} = k^2 \left[\hat{\phi}({\bf k},t) - k^2 \hat{\phi}({\bf k},t) - \sum_{{\bf k}_1,{\bf k}_2,{\bf k}_3}^{'} \hat{\phi}({\bf k}_1,t) \hat{\phi}({\bf k}_2,t) \hat{\phi}({\bf k}_3,t) \right] ,
\label{chc_kspace}
\end{eqnarray} 
where $\hat{\phi}({\bf k},t) = \sum_{{\bf k}} e^{i {\bf k} \cdot {\bf r}}\phi({\bf r},t)$, and the prime denotes ${\bf k}={\bf k}_1+{\bf k}_2+{\bf k}_3$. We multiply Eq.~(\ref{chc_kspace}) by $\phi^{*}({\bf k}, t )$, and add the resultant equation to its complex conjugate to obtain the following equation for the spectral energy: 
\begin{eqnarray}
 \frac{\partial E({\bf k},t)}{\partial t}  =  2k^2E({\bf k},t) - 2k^4 E({\bf k},t) + k^2T({\bf k},t).
\label{chc_spectral}
\end{eqnarray}
In Eq.~(\ref{chc_spectral}), $E({\bf k},t) = |\hat{\phi}({\bf k},t)|^2 / 2$ is the modal energy. The terms on the RHS of Eq.~(\ref{chc_spectral}) have the following significance: (a) the term $2k^2 E({\bf k},t)$ feeds energy into the wavenumber ${\bf k}$; (b) the $2k^4 E({\bf k},t)$ term dissipates energy; and (c) $k^2T({\bf k},t)$ is the nonlinear energy transfer rate~\cite{greenwade93,verma2019energy,verma2021variable}:
\begin{equation}
k^2T({\bf k},t)  =   
-k^2\sum_{{\bf k}_1,{\bf k}_2,{\bf k}_3}^{'} \operatorname{Re}\left\{\hat{\phi}({\bf k}_1,t) \hat{\phi}({\bf k}_2,t) \hat{\phi}({\bf k}_3,t) \hat{\phi}^*({\bf k},t) \right\}.
\label{chc_tkt}
\end{equation}
  
Since $\phi({\bf r},t)$ is a real function, $\phi({\bf -k},t) = \phi^*({\bf k},t)$. Therefore, $E({\bf -k},t) = E({\bf k},t)$. Following the turbulence literature, we also define the $1D$ energy spectrum $E(k,t)$, which is a sum of modal energies over a shell of unit width:
\be
E(k,t) = \sum_{k-1 < k' \le k} E({\bf k'}, t).
\ee
For 1D, $E(k,t) = 2E({\bf k},t)$ for $k>0$, and $E(k=0,t) = E({\bf k}=0,t)$. For 2D, the modal energies $E({\bf k},t)$ are independent of the angle due to isotropy. Therefore, $E(k,t) \simeq 2\pi k E({\bf k},t)$. In this paper, we study the total modal energy of the system ${\tilde E}(t) = \sum_{{\bf k}} |\phi({\bf k}, t )|^2/2 = \int d{\bf r}~\phi({\bf r},t)^2/2$ rather than the usual GL free energy $\mathcal{F}[\phi]$.

 Note that the evolution of the CH equation towards the asymptotic state ($\phi^{*} = \pm 1$) is contingent on the initial conditions. We consider cases with: (i) critical $\phi({\bf r},0)$, where $\phi_0 = 0$; (ii) biased $\phi({\bf r},0)$, where $\phi_{0} \ne 0$. 
 
 \subsubsection{Critical Case} \label{Methodology_II_cahn_hilliard_critical}
 
 In this case, the initial condition consists of small fluctuations about $\phi_0 = 0$. We consider the CH equation in Eq.~(\ref{chcspace}). Initially, the linear term dominates compared to the nonlinear term. Consequently, the evolution is primarily governed by the difference between energy injection and linear energy dissipation. Thus, the energy spectrum evolves as 
\begin{eqnarray}
\frac{\partial E({\bf k},t)}{\partial t} \simeq 2k^2(1-k^{2})E ({\bf k}, t) ,
\label{Ek_max_exp}
\end{eqnarray}
with the solution
\begin{eqnarray}
E({\bf k},t) \simeq e^{2k^2(1-k^2)t} E({\bf k},0) .
\label{Ek_max_exp1_solution}
\end{eqnarray}
Thus, all modes with $0< k <1$ grow exponentially, with the fastest growth being for $k_m = 1/\sqrt{2}$.  In this early regime, both $ \hat{\phi}({\bf k},t)$ and ${\tilde E}(t)$ show significant growth in time. As the order parameter amplitude grows and domain growth is initiated, nonlinear energy transfer prevails over linear energy dissipation. This transition leads to the formation and merger of kink-antikink pairs or domains in $D = 1$. In the bulk of the domain, $\phi$ takes values $\pm 1$. The value of $\phi$ differs significantly from $\pm 1$ only on domain walls -- $\phi = 0$ defines the location of the interface. In $D = 2$, the domain's interior is thermalized in one of the equilibrium phases. The domain boundaries slowly move and tend to reduce curvature due to their elastic energy. In the bulk of the domain,
\be
\phi^3({\bf r},t) = \phi^2({\bf r},t) \phi ({\bf r},t) \simeq \phi({\bf r},t).
\ee
 Therefore, from Eq.~(\ref{chc_tkt}) 
\bea
{ k}^2T({\bf k},t) \simeq - { k}^2\phi({\bf k},t) \phi^*({\bf k},t)  = -2 {k}^2 E({\bf k},t).
\label{chc_Tk_2ek}
\eea
In terms of shell variables,
\be
k^2 T(k, t) \simeq -2 k^2 E(k,t).
\ee
Thus, the nonlinear term induces dissipation in the bulk, which plays a critical role in the coarsening process.

During the approach to the asymptotic state, $\partial_t E({ k},t) \simeq 0$, implying that ${k}^2 T(k,t) \simeq 2{k}^2(1-k^2)E(k,t)$. At small wavenumbers ($k \ll 1$), the dissipation term $2 k^4 E(k,t)$ is much smaller than the energy supply rate $2 k^2 E(k,t)$. Therefore, the nonlinear term $-k^2 T(k,t)$ must balance $2 k^2 E(k,t)$ at small $k$. 

\subsubsection{Biased Case}
\label{MethodologyIIcahnhilliardbiased}

Next, we consider the biased case with $\phi_{0} \ne 0$.  The linear stability analysis of Eq.~(\ref{chcspace}) reveals that a random initial condition with $|\phi_0|< 1/\sqrt{3}$ is unstable for small $k$. After the quench, $\phi ({\bf r}, t)$ evolves towards its fixed point values, $\phi^{*} = \pm 1$. This evolution of $\phi({\bf r},t)$ closely parallels the behavior for $\phi_{0}=0$ in the coarsening and asymptotic regime. The system forms kinks and antikinks, which annihilate on collision. Therefore, the scenario discussed above in Sec.~\ref{Methodology_II_cahn_hilliard_critical} applies again. The asymptotic state of the system consists of two large domains (of $+1$ and $-1$, respectively). The relative domain sizes are such that the average order parameter is $\phi_{0}$.

In the case with $|\phi_{0}| > 1/\sqrt{3}$, the initial fluctuations are exponentially damped and the asymptotic state is $\phi({\bf r}, t) = \phi_{0}$. In this case, we only need to consider the linearized CH equation:
\begin{eqnarray}
\frac{{\partial \delta\phi({\bf r},t)}}{\partial t} &=& -\nabla^2\left[(1-3\phi_0^2)\delta \phi +  \nabla^2\delta\phi\right],
\label{chclinearization}
\end{eqnarray}
where $\phi({\bf r},t) = \phi_0 + \delta\phi({\bf r},t)$. In Fourier space, the fluctuations obey
\begin{eqnarray}
\frac{{\partial \delta \hat{\phi}({\bf k},t)}}{\partial t} &\simeq& -k^2\left[(3\phi_0^2-1+k^2)\delta{\hat\phi}\right] 
\label{chclinear}
\end{eqnarray}
with the solution 
\begin{eqnarray}
\delta \hat{\phi}({\bf k},t) &=& e^{-k^2(3\phi_0^2-1+k^2)t}\delta \hat{\phi}({\bf k},0).
\label{chclinearizationkspacesolution}
\end{eqnarray}

With the above solution, it is straightforward to obtain $E({\bf k},t)$ and $T({\bf k},t)$. Thus, we have
\begin{eqnarray}   
E({\bf k},t) &=& \frac{|\hat{\phi}({\bf k},t)|^2}{2} \nonumber \\ 
&=& \frac{\phi_0^2}{2}\delta({\bf k})^2 + \frac{1}{2}|\delta\hat{\phi}({\bf k})|^2.
\end{eqnarray}
We have dropped the linear terms in $\delta\hat{\phi}$ as these variables are Gaussian with zero average. Thus, an ensemble averaging eliminates the linear terms. Further, the energy transfer $T({\bf k}, t)$ arises from the linearized term $-3\phi_0^2\delta{\hat\phi}$ in Eq.~(\ref{chclinear}). Thus
\begin{eqnarray}
T({\bf k}, t) &=& -3\phi_0^2|\delta\hat{\phi}({\bf k}, t)|^2 \nonumber \\ 
&=& -6\phi_0^2 E({\bf k}, t) \quad \text{for} \ {\bf k} \ne 0. 
\label{tk_6ek}
\end{eqnarray}

\subsection{Time-dependent Ginzburg-Landau Equation}\label{MethodologyIITDGL}
 
The next system we consider is a ferromagnet consisting of atomic spins $\{S_{i}\}$ which take values $S_{i} =\pm 1$. If the system is quenched from the paramagnetic state to the ferromagnetic state at $t=0$, it evolves via the emergence and growth of domains enriched in the  ``up'' and ``down'' phases \cite{pw09}. The evolution is characterized by a nonconserved order parameter (NCOP). At the coarse-grained level, this NCOP dynamics obeys the TDGL equation:
 \begin{eqnarray}
\frac{\partial \phi({\bf r},t) }{\partial t}  &=&- \frac{\delta \mathcal{F}[\phi]}{\delta \phi} , 
\label{tdgl_fun}
\end{eqnarray}
where $\phi({\bf r},t) $ is the local magnetization. We substitute $\mathcal{F}[\phi]$ from Eq.~(\ref{functional_energy}) in Eq.~(\ref{tdgl_fun}) to obtain
\begin{eqnarray}
\frac{{\partial \phi({\bf r},t)}}{\partial t} &=&  \phi- \phi^3 + \nabla^2\phi.
\label{tdgl_space}   
\end{eqnarray}

In Fourier space, the TDGL equation (\ref{tdgl_space}) takes the form 
\begin{eqnarray}
\frac{\partial \hat{\phi}({\bf k},t)}{\partial t} = \hat{\phi}({\bf k},t) - k^2 \hat{\phi}({\bf k},t) - \sum_{{\bf k}_1,{\bf k}_2,{\bf k}_3}^{'} \hat{\phi}({\bf k}_1,t) \hat{\phi}({\bf k}_2,t) \hat{\phi}({\bf k}_3,t) . 
\label{tdgl_kspace}
\end{eqnarray} 
In Ref.~\cite{verma2023nonlinear}, we studied the spectral energy transfer properties of Eq.~(\ref{tdgl_kspace}) in $D=1$ following the same procedure as outlined in Sec.~\ref{Methodology_II_cahn_hilliard}. Here, we summarize our earlier results. An initial condition  with $\phi_{0}=0$ evolves to a sequence of kink-antikink pairs that asymptote to either $\phi(x,t) = 1$ or $\phi(x,t) = -1$. During the approach to the asymptotic state, $T(k,t) \simeq - 2 E(k,t)$ for small $k$.

On the other hand, when the initial condition is random with a uniformly positive (or uniformly negative) $\phi(x,0)$, the system evolves towards the asymptotic state where $\phi(x,t) = 1$ ($\phi(x,t) = -1$). This evolution can be studied in a linear approximation. For arbitrary $D$, we decompose $\phi({\bf r},t)$ and $\hat{\phi}({\bf k},t)$ as 
\begin{eqnarray}
\phi({\bf r},t) &=& a(t) + \delta\phi({\bf r},t),  \nonumber  \\ 
\hat{\phi}({\bf k},t) &=& a(t)\delta({\bf k}) + \delta{\hat\phi}({\bf k},t) ,
\end{eqnarray}
where $a(t)$ is the time-dependent average value of $\phi$. These quantities obey the evolution equations:
\begin{eqnarray}
\frac{da(t)}{dt} &=& a-a^3, 
\label{tdgl_in_a(t)} \\  
\frac{\partial \delta \phi({\bf r},t)}{\partial t} &=& (1-3a^2 + \nabla^2)\delta \phi .
\label{tdgl_in_deltaphi(t)}
\end{eqnarray}
The linearization in Eq.~(\ref{tdgl_in_deltaphi(t)}) is valid for $a(t) > 1/\sqrt{3}$. The solutions for Eq.~(\ref{tdgl_in_a(t)}) are $a^* = 0$ (unstable), and
\begin{eqnarray}
a(t) &=& \frac{\phi_0}{\sqrt{\phi_0^2 + (1-\phi_0^2)e^{-2t}}},
\end{eqnarray}
so that $a(t) \to $ sgn($\phi_{0}$) as $t \to \infty$.

The solution of Eq.~(\ref{tdgl_in_deltaphi(t)}) in Fourier space is 
\begin{eqnarray}
\delta \hat{\phi}({\bf k},t) &=& \exp \left[ -\int_{0}^{t} d\tau \left(3a^2 - 1 + k^2 \right) \right] \delta\hat\phi({\bf k},0).
\end{eqnarray}

Unlike the CH equation, the average value of $\phi({\bf r},t)$ is time-dependent as there is no conservation constraint. It rapidly settles to the asymptotic value $a(t) = \pm 1$. It is straightforward to obtain the corresponding linearized expressions for the modal energy and energy transfer:
\begin{eqnarray}
E({\bf k},t ) &=& \frac{a(t)^2}{2}\delta({\bf k})^2 + \frac{|\delta {\hat\phi}({\bf k},t)|^2}{2}, \\ \nonumber 
T({\bf k},t ) &=& -3a(t)^2|\delta {\hat\phi}({\bf k},t)|^2 \nonumber \\
&=& -6a(t)^2E({\bf k},t) \quad \text{for} \ {\bf k} \ne 0.
\label{Tkt}
\end{eqnarray}

During this evolution, we can treat $\hat{\phi} ({\bf k}, t )$ as Gaussian to obtain $T ({\bf k},t) \simeq -12 \allowbreak \tilde{E}(t) E({\bf k},t)$. In the asymptotic regime, $\tilde{E}(t) \simeq 1/2$; therefore, $T({\bf k},t) \allowbreak \simeq -6E({\bf k},t)$ in accordance with Eq.~(\ref{Tkt}). In this paper, for completeness, we extend our earlier investigation of the TDGL equation to $D=2$.

\section{Details of Numerical Simulations}
\label{s3}

In our current study, we performed simulations of the CH and TDGL equations in $D=1,2$ using the finite-difference method. We took a system with linear dimensions denoted as $L$, and applied periodic boundary conditions in all directions. The specific details regarding grid resolution $(N)$ and space mesh size ($\Delta x$), time step size ($\Delta t$), and the system size ($L$) are provided in Table~\ref{sim_params}.

\begin{table}
\centering
\caption{Simulation Parameters for the CH and TDGL Equations}
\begin{ruledtabular}
\begin{tabular}{ccccccc}
Model & $D$ & $L$ & $N$ & $\Delta x = L/N$ & $\Delta t$ \\
\hline
CH & 1 & 100 & 512 & 0.195 & $10^{-4}$ \\
TDGL & 1 & 100 & 1024 & 0.097 & $10^{-3}$ \\
CH & 2 & 256 & 256 & 1 & $10^{-2}$ \\
TDGL & 2 & 128 & 128 & 1& $10^{-2}$ \\
\end{tabular}
\end{ruledtabular}
\label{sim_params}
\end{table}

To perform these simulations, we utilized the central-difference scheme for the spatial Laplacian, and the forward Euler scheme for time differencing:
\bea
\label{Lderv}
\frac{\partial^2 \phi(x,t)}{\partial x^2} &=& \frac{\phi(x+\Delta x,t) - 2\phi(x,t) + \phi(x-\Delta x,t)}{\Delta x^2} , \\
\label{derv}
\frac{\partial \phi(x,t)}{\partial t} &=& \frac{\phi(x,t+ \Delta t) -\phi(x,t)} {\Delta t} .
\eea
The mesh sizes $\Delta x$ and $\Delta t$ were chosen to ensure compliance with the numerical stability criteria. These are derived from a linear stability analysis of the CH and TDGL equations. We choose the grid spacing $\Delta x = L/N$ in $D=1,2$. This has to be small enough to capture the interface, which is of width $\xi = \sqrt{2}$ in our dimensionless units.

The time step $\Delta t$ satisfies a {\it stability criterion}. This is obtained by the requirement that the discrete dynamical system should not exhibit an unphysical subharmonic bifurcation about the fixed points $\phi^* = \pm 1$ \cite{santra2011computational}. For the CH equation, we ensured that the time step satisfies the following stability condition:
\begin{eqnarray}
\Delta t \le \frac{(\Delta x)^4}{4D [(\Delta x)^{2} + 2D]} .
\end{eqnarray}
Similarly, for the TDGL equation, we adhered to the stability condition:
\begin{eqnarray}
\Delta t \le \frac {(\Delta x)^2}{(\Delta x)^{2} + 2D}.
\end{eqnarray}

\section{Numerical Results for the Cahn-Hilliard Equation}
\label{s4}

In this section, we present numerical results for the CH equation in $D = 1, 2$. We consider two distinct classes of initial conditions as discussed in Sec.~\ref{Methodology_II_cahn_hilliard}. In particular, we analyze the energy spectrum and energy transfers in the coarsening regime. 

\subsection{Energy Transfers in the Cahn-Hilliard Equation with \texorpdfstring{$\phi_{0} =0$}{Lg}}
\label{Numerical1}

\begin{figure}[!ht]
\centering
\includegraphics[scale=0.7]{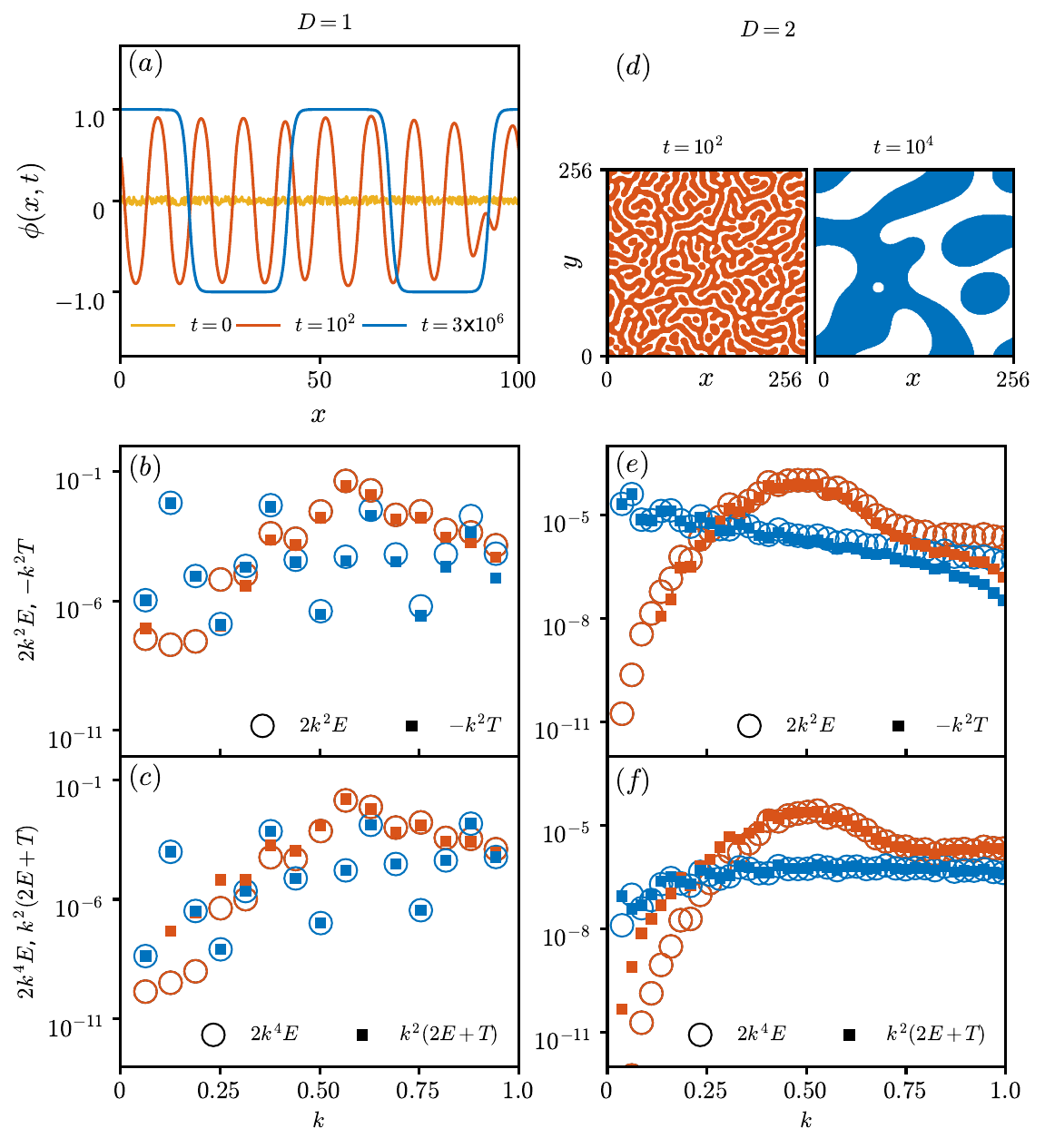}
\caption{Simulation of the CH equation with $\phi_0 = 0$ in $D=1,2$. (a) Order parameter profiles in $1D$ at $t=0,10^2,3 \times 10^6$. (b) Linear-log plot of $2k^2 E(k,t)$ and $-k^2 T(k,t)$ vs. $k$ at $t=10^2,3 \times 10^6$ in $1D$. (c) Linear-log plot of $k^2 (2E(k,t)+T(k,t)))$ and $2k^4 E(k,t)$ vs. $k$ at same times as in (b). (d) Snapshots of the order parameter in $2D$ at $t = 10^2, 10^4$. The colored and uncolored regions denote $\phi > 0$ and $\phi < 0$, respectively. (e) Linear-log plot of $2k^2E(k,t)$ and $-k^2 T(k,t)$ vs. $k$ at $t = 10^2, 10^4$ in $2D$. (f) Linear-log plot of $k^2 (2E(k,t)+T(k,t))$ and $2k^4E(k,t)$ vs. $k$ at same times as in (e).}
\label{fig1}
\end{figure}

Let us start by discussing the evolution of $\phi({\bf r}, t)$ in Eq.~(\ref{chcspace}) when $\phi_{0} =0$. Fig.~\ref{fig1}(a) shows the evolution of $\phi$ in $D=1$ at times $t = 0$, $10^2$, $3 \times 10^6$. Initially, fluctuations in $\phi$ grow, followed by their saturation and the formation of kink-antikink pairs.

In Fig.~\ref{fig1}(b), motivated by Eq.~(\ref{chc_Tk_2ek}), we plot $2k^2 E(k,t)$ and $−k^2T(k,t)$ vs. $k$. The color-coding in Fig.~\ref{fig1}(b) corresponds to the same times as in Fig.~\ref{fig1}(a). It is important to note that, at $t=10^2$, $\phi$ differs substantially from $\pm 1$, leading to a marked deviation between $2k^2 E(k,t)$ and $−k^2T(k,t)$. There are even some small $k$-values where $-k^2T(k,t)$ is negative -- these are not shown in the linear-log plot of Fig.~\ref{fig1}(b). This difference kills the fluctuations and contributes to the formation of kink-antikink structures. However, at $t=3 \times 10^6$, $\phi \simeq \pm 1$ throughout the system. Consequently, we observe $2k^2E(k,t) \simeq −k^2T(k,t)$ for small $k$. Any discrepancies between these quantities for larger $k$ can be attributed to terms involving $2k^4E(k,t)$.

In Fig.~\ref{fig1}(c), we have plotted $k^2 (2 E(k,t) + T(k,t))$ and $2k^4 E(k,t)$ vs. $k$. At $t=10^2$, for $k \ll 1$, $k^2 (2 E(k,t) + T(k,t))$ and $2k^4 E(k,t)$ are not equal, and the difference is the reason $\phi$ evolves with time. For $t = 3 \times 10^6$, $k^2 (2 E(k,t) + T(k,t))$ and $2k^4 E(k,t)$ are approximately equal, implying that $\partial_{t}E(k,t) \simeq 0$. Therefore, $\phi$ evolves extremely slowly.

Next, we consider the CH equation with $\phi_{0} = 0$ in $D=2$. The evolution snapshots at $t=10^2$ and $10^4$ are shown in Fig.~\ref{fig1}(d). The colored regions represent areas with $\phi > 0$, while the white regions represent $\phi<0$. Small domains tend to collapse and merge into larger domains to minimize free energy. The order parameter is uniform in the bulk region of domains ($\phi = \pm 1$). The excess free energy is stored in the domain walls. 

In Fig.~\ref{fig1}(e), we plot $2k^2E(k,t)$ and $−k^2T(k,t)$ vs. $k$ for the snapshots in Fig.~\ref{fig1}(d). In the coarsening regime, the difference between these two quantities (especially for small values of $k$) drives the coarsening process. At later time $(t=10^4)$, the difference becomes significant only at larger values of $k$, where the term $2k^4 E(k,t)$ also plays a role. In Fig.~\ref{fig1}(f), we plot
$k^2 (2 E(k,t) + T(k,t))$ and $2k^4 E(k,t)$ vs. $k$ for the same profiles. At late times $(t=10^4)$, these quantities are almost coincident as the dynamics of $\phi$ becomes extremely slow as phase separation proceeds.

\subsection{Energy Transfers Between Modes During Coarsening}\label{Numerical_Results_Ek_tk_in_modes}

\begin{figure}[ht]
\centering
\includegraphics[scale=0.6]{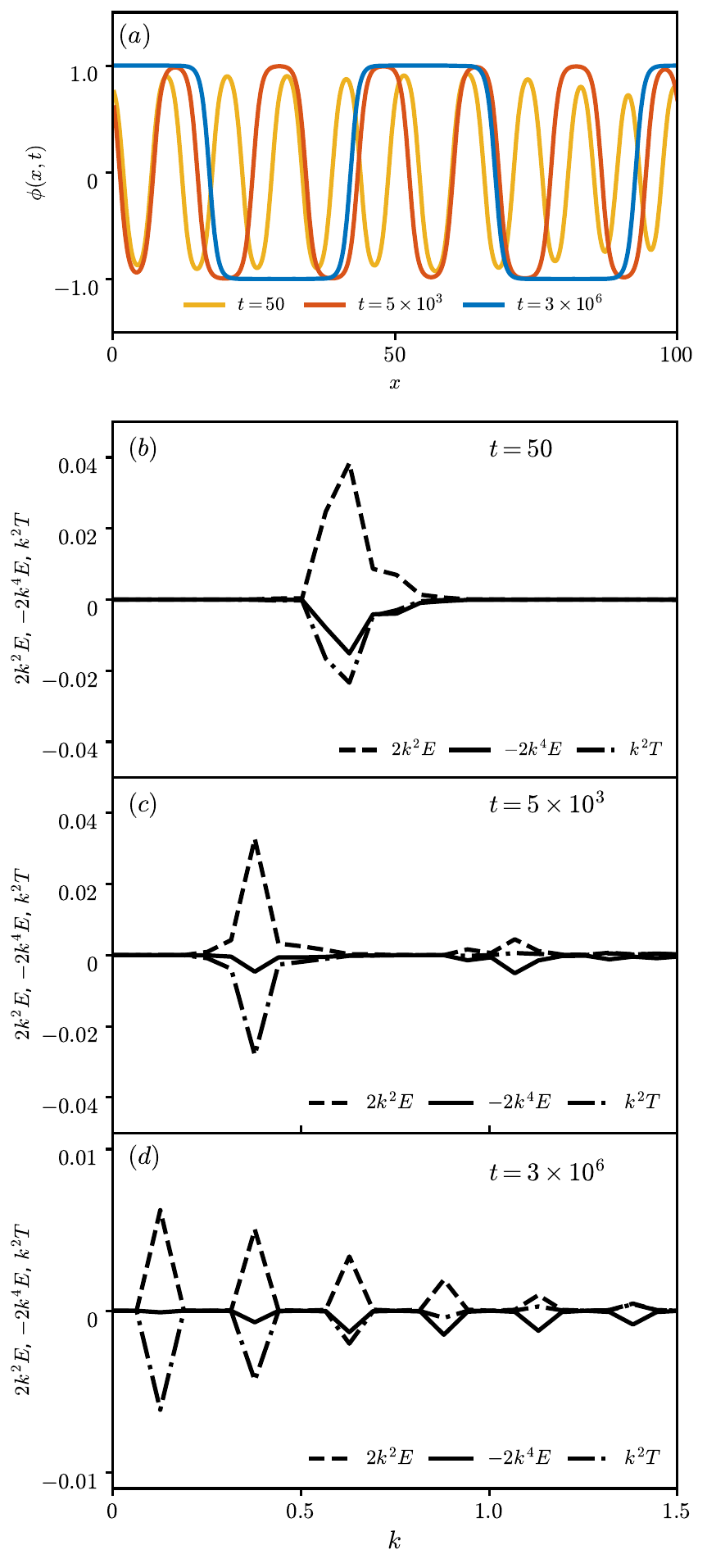}
\caption{(a) Order parameter profiles in $1D$ at $t=50, 5 \times 10^3,  3 \times 10^6$. (b)-(d) show the spectral representation of $2k^2E(k,t),-2k^4E(k,t)$ and $k^2T(k,t)$ vs. $k$ for the three times in (a).}
\label{fig2}
\end{figure}

We examine mode-to-mode energy transfers in the $1D$ CH equation for $\phi_{0}=0$, revealing dominant balances among different scales. The three terms on the RHS of Eq.~(\ref{chc_spectral}) are obtained from a numerical solution of Eq.~(\ref{chcspace}). The evolution snapshots at $3$ times are shown in Fig.~\ref{fig2}(a). The corresponding terms from  Eq.~(\ref{chc_spectral}) are plotted in Fig.~\ref{fig2}(b)-(d).

In the coarsening regime, energy injection and dissipation peak at the same $k$ but with varying magnitudes. Notably, the linear energy dissipation rate $2k^4 E(k,t)$ is significantly smaller than $ -k^2T(k,t)$. Consequently, $k^2 T(k,t)$ is the prime destroyer of fluctuations for small $k$. However, at larger wavenumbers $(k > 1)$, the dissipation term $2k^4E(k,t)$ becomes dominant due to the prefactor.

In the asymptotic state, where $\phi(x,t)$ settles to $\pm 1$ throughout the system, $-k^2T(k,t) \simeq 2k^2 E(k,t)$ for $k < 1$, as shown in Fig.~\ref{fig2}(d). Notably, energy is primarily concentrated in modes with $k \ll 1$. Further, $E(k = 0)$ remains constant throughout the evolution of $\phi$, due to the conservation law.

\subsection{Energy Spectrum in the Linear Regime}
\label{Numerical_Results_Sub3}

\begin{figure}[ht]
\centering
\includegraphics[scale=0.7]{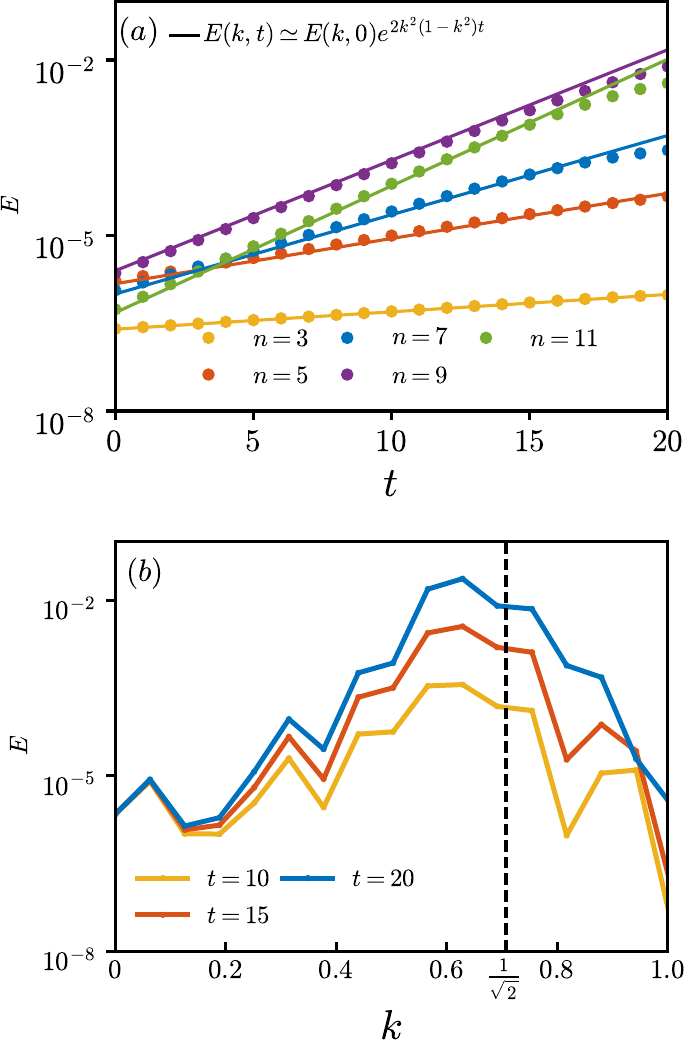}
\caption{(a) Plot of $E(k,t)$ vs. $t$ at early times for different $k$ for $\phi_0 = 0$. The values of $k$ are $2 \pi n/L$, where the mode number is specified in the key. The lines denote the functional form $E(k,t)= E(k,0) \exp [2k^2 (1-k^2)t]$. (b) Plot of $E(k,t)$ vs. $k$ for $t = 10,15,20$. The most unstable wave-vector, $k_m = 1/\sqrt{2}$, is marked by a dashed line.}
\label{fig3}
\end{figure}

In accordance with the discussion in Sec.~\ref{s1}, during the initial stages, the evolution of $\phi$ does not depend on the nonlinear term. In this linear regime, we have 
\bea
\hat{\phi}({\bf k},t) &\simeq& \hat{\phi}({\bf k},0)e^{k^2(1-k^2)t} , \\
E({\bf k},t) &\simeq& E({\bf k},0)e^{2k^2(1-k^2)t}.
\eea
We calculate $E(k,t)$ numerically and fit the data with the above expression in Fig.~\ref{fig3}(a). Notably, the energy spectrum peaks around $k \simeq 1/\sqrt{2}$, as shown in Fig.~\ref{fig3}(b).      

\subsection{Energy Spectrum \texorpdfstring{$E(k,t)$}{Lg} and Total Energy \texorpdfstring{$\tilde{E}(t)$}{Lg}}
\label{NumericalResultsSub4}

\begin{figure}[ht]
\centering
\includegraphics[scale=0.80]{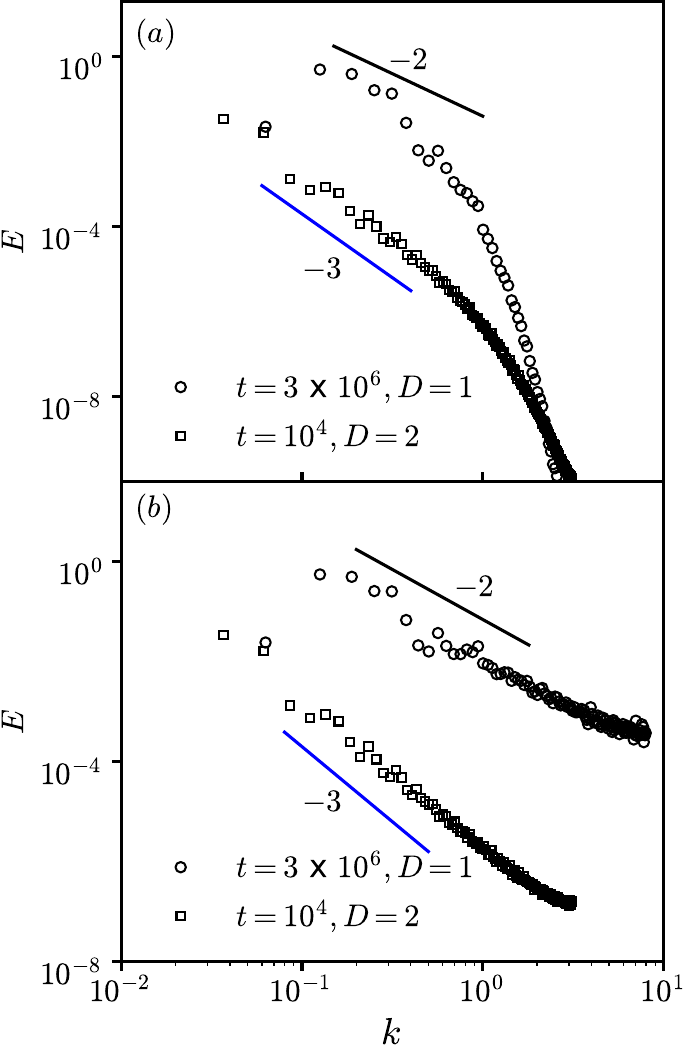}
\caption{(a) Plot of energy spectra, $E(k,t)$ vs. $k$, for $D=1$ at $t=3 \times 10^6$; and for $D=2$ at $t=10^4$. The black line indicates Porod's law [$E(k,t) \sim k^{-2}$] in $D=1$, whereas the blue line indicates Porod's law [$E(k,t) \sim k^{−3}$] in $D=2$. (b) Analogous to (a), but for the hardened order parameter fields.}
\label{fig4}
\end{figure}

The plots in Fig.~\ref{fig4}(a) show the energy spectrum $E(k,t)$ vs. $k$ for the snapshots of $\phi({\bf r}, t)$ in Fig.~\ref{fig1}(a) at $t = 3 \times 10^6$ $(D=1)$; and  Fig.~\ref{fig1}(d) at $t=10^4$ $(D=2)$. Due to the conservation law, $E({\bf k}=0,t)=0$ and the spectrum rises to a peak at $k_m(t)$. In Fig~\ref{fig4}, we show only data for $k > k_m$. The steep variations of $\phi$ at the domain boundaries show a power law scaling: $E(k,t)\sim k^{-2}$ in $D=1$, and $E(k,t) \sim k^{-3}$ in $D=2$. This is known as the {\it Porod law} \cite{porod1982small} in the domain growth literature. The modal energy is analogous to the {\it structure factor} in coarsening studies \cite{pw09}, and Porod's law is a consequence of scattering from sharp interfaces. The deviation from Porod's law seen at values of $k>\xi^{-1}$ (where $\xi = \sqrt {2}$ is the kink width) is due to the softness of the interface \cite{OonoPURI_1988}.

We can extend the Porod window by ``hardening'' the order parameter profiles~\cite{op86}. In this procedure, we assign $\phi(\mathbf{r}, t) = 1$ for all points where $\phi(\mathbf{r},t) > 0$,  and $\phi(\mathbf{r},t) = -1$ for all points where $\phi(\mathbf{r}, t) < 0$. After this hardening, we again calculate the energy spectra, which are shown in Fig.~\ref{fig4}(b). The Porod law is also observed in Burgers turbulence, where $\phi({\bf r}, t)$ exhibits sharp jumps at the shocks ~\cite{frisch1995turbulence,verma2000intermittency}. 

\begin{figure}[ht]
\centering
\includegraphics[scale=0.80]{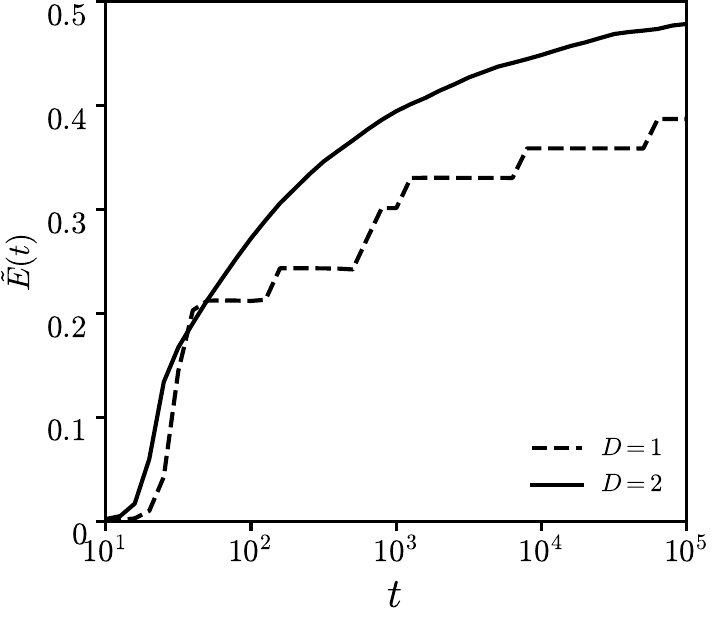}
\caption{Plot of $\tilde{E}(t)$ vs. $t$ for the CH equation in $D=1,2$ for $\phi_0 = 0$.}
\label{fig5}
\end{figure}

In Fig.~\ref{fig5}, we plot the total energy $\tilde {E}(t)$ vs. $t$ for the evolution of $\phi$ in $D=1, 2$ for $\phi_{0}=0$. The figure shows a gradual increase in $\tilde{E}(t)$ over time, reaching saturation at $\tilde {E} = 1/2$ . The merger of kink-antikink pairs induces jumps, as the interfacial area (where $\phi \simeq 0$) is reduced. This process is particularly evident for the $D=1$ case, as the kink-antikink annihilation is a sudden process.

\subsection{Energy Transfers with Biased Initial Conditions (\texorpdfstring{$\phi_{0} \ne 0$}{Lg})}
\label{NumericalResultschcbiased}

Our results so far have focused on the spectral energy transfers in the CH equation during the evolution of $\phi$ for $\phi_{0} =0$.  Now, we turn our attention to the case of biased initial conditions, $\phi_{0} \ne 0$. In Sec.~\ref{MethodologyIIcahnhilliardbiased}, we observed that the system forms kink-antikink pairs for $|\phi_{0}| <1/\sqrt{3}$. Then, the evolution is analogous to the critical case, and the energy transfer scenario is the same as that in Sec.~\ref{Numerical1}.

On the other hand, for $|\phi_{0}| >1/\sqrt{3}$, the initial fluctuations are damped exponentially and the asymptotic solution is $\phi({\bf r}, t)\simeq \phi_{0}$. In that case, the relevant energy transfers obey Eq.~(\ref{tk_6ek}). As this case is analytically tractable, we do not show the corresponding numerical results here.

\section{Numerical Results for the TDGL Equation}
\label{s5}

In our previous paper~\cite{verma2023nonlinear}, we have discussed energy transfers in the $D=1$ TDGL equation. For completeness, in this section. we briefly present numerical results for the $D=2$ TDGL equation, and compare them with the $D=1$ case. The theoretical framework for the TDGL equation is presented in Ref.~\cite{verma2023nonlinear}, and summarized in  Sec.~\ref{MethodologyIITDGL} of this paper.

\subsection{Energy Transfers in the TDGL Equation with \texorpdfstring{$\phi_{0} = 0$}{Lg}}

\begin{figure}[ht]
\centering
\includegraphics[scale=0.7]{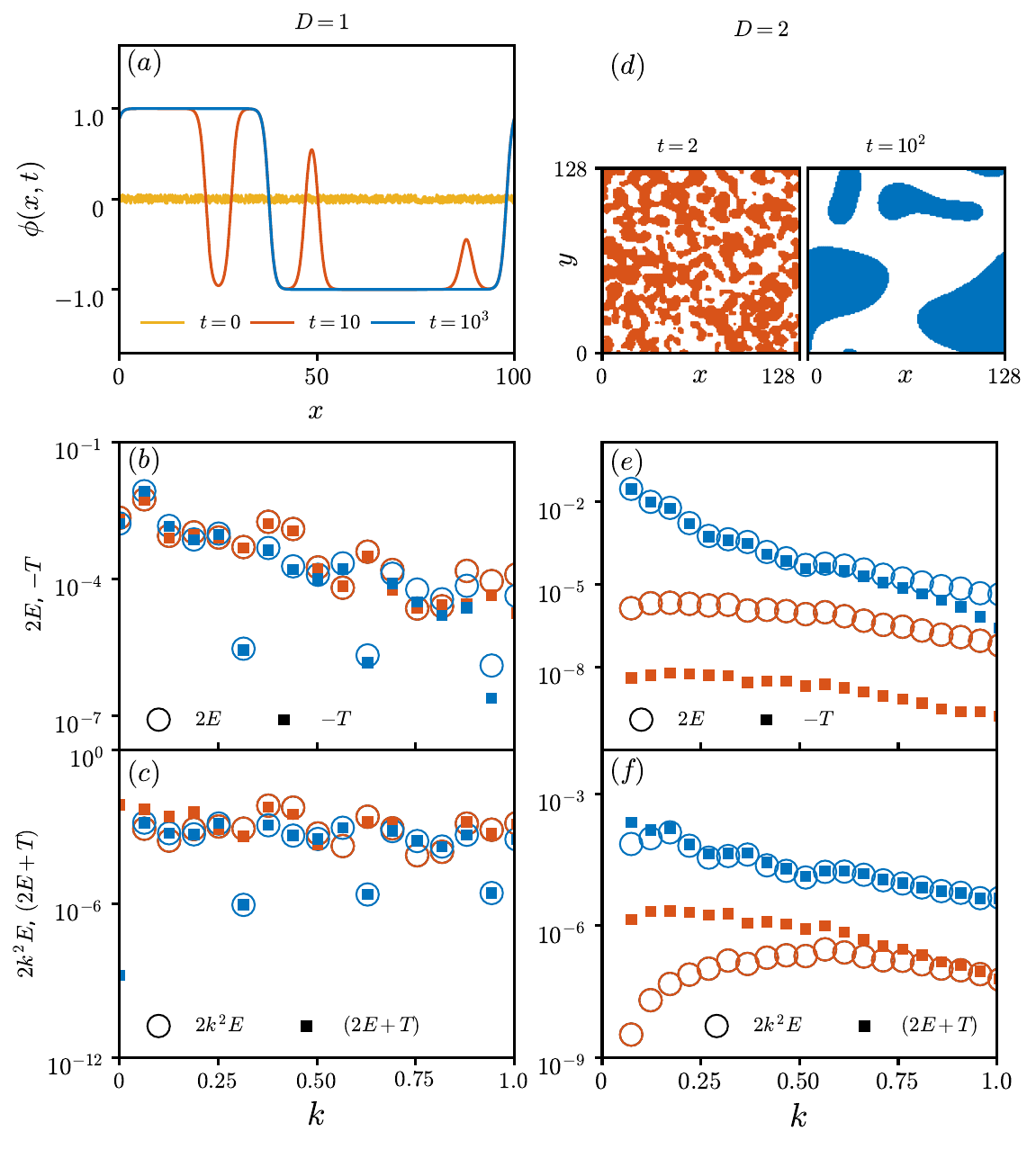}
\caption{Simulation of the TDGL equation with $\phi_0 = 0$ in $D=1,2$.
(a) Order parameter profiles in $1D$ at $t=0,10,10^3$. (b) Plot of $2E(k,t)$ and $-T(k,t)$ vs. $k$ in $1D$ at $t=10,10^3$. (c) Plot of $2k^2 E(k,t)$ and $(2E(k,t)+T(k,t))$ vs. $k$ for the same times as (b). (d) Snapshots of the order parameter in $2D$ at $t = 2,10^2$. The colored and uncolored regions denote $\phi > 0$ and $\phi < 0$, respectively. (e) Plot of $2E(k,t)$ and $-T(k,t)$ vs. $k$ in $2D$ at $t = 2,10^2$. (f) Plot of $2k^2 E(k,t)$ and $(2E(k,t) + T(k,t))$ vs. $k$ for the same times as (e).}
\label{fig6}
\end{figure}

Let us first present numerical results for the TDGL equation with a critical composition. We initialize the system with random small fluctuations $\delta \phi ({\bf r},0) \in [-0.05,0.05]$ around $\phi_{0} = 0$, representing the pre-quench paramagnetic phase. First, we recall our earlier results \cite{verma2023nonlinear}. The 1D evolution of $\phi(x,t)$ is illustrated in Fig.~\ref{fig6}(a) at $t = 0$, $10$, and $10^3$. As time progresses, fluctuations in $\phi$ diminish and $\phi \to \pm 1$ in the asymptotic state, resulting in a total energy ${\tilde E}(t) = 1/2$. In Fig.~\ref{fig6}(b), we plot $2E(k,t)$ and $-T(k,t)$ vs. $k$ for the profiles at $t=10, 10^3$. At $t=10^3$, both quantities show a nice data collapse for small $k$, consistent with the discussion in Sec.~\ref{MethodologyIITDGL}. In Fig.~\ref{fig6}(c), we plot $T(k,t)+2E(k,t)$ and $2k^2E(k,t)$ vs. $k$. The scales for these quantities are very small, in general. At $t = 10$, we observe that $T(k,t) + 2E(k,t)$ differs substantially from $2k^2E(k,t)$ for small $k$, indicating that $\partial_t E(k,t)$ is significant.

Next, we present the numerical results for $D=2$. We show evolution snapshots of $\phi({\bf r}, t)$ at $t = 2$, $100$ in Fig.~\ref{fig6}(d). The white regions represent $\phi < 0$, whereas colored regions correspond to $\phi > 0$. In Fig.~\ref{fig6}(e), we plot $2E(k,t)$ and $-T(k,t)$ vs. $k$. The discrepancy between these quantities for $k < 1$ is significant at $t = 2$, where $\phi^3({\bf r},t) \ne \phi({\bf r},t)$. However, by $t = 10^2$, both quantities exhibit a pronounced data collapse at small $k$, aligning with our discussion in Sec.~\ref{MethodologyIITDGL}. We plot $T(k,t) + 2E(k,t)$ and $2k^2E(k,t)$ vs. $k$ in Fig.~\ref{fig6}(f). For small $k$, $T(k,t) + 2E(k,t)$ and $2k^2 E(k,t)$ differ substantially at $t=2$, hence $\partial_t E(k,t) \neq 0$.  These results are comparable to our $D=1$ results. 

\subsection{Energy Spectrum \texorpdfstring{$E(k,t)$}{Lg}}

\begin{figure}[ht]
\centering
\includegraphics[scale=0.70]{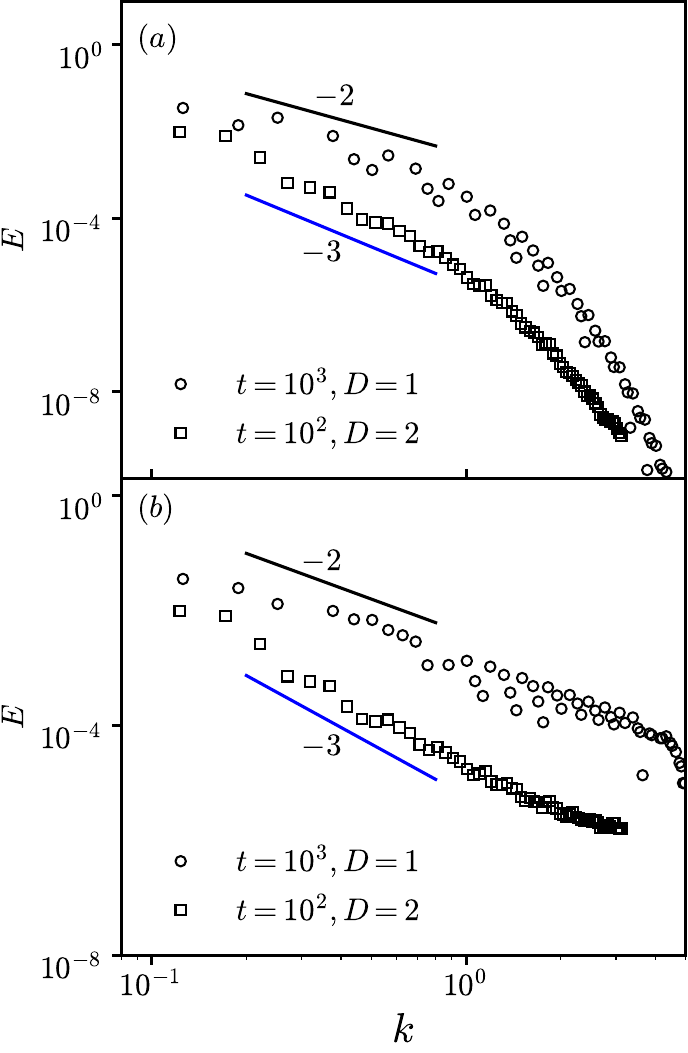}
\caption{Analogous to Fig.~\ref{fig4}, but for the TDGL equation in $D=1,2$.}
\label{fig7}
\end{figure}

Next, we study the energy spectra, specifically examining the behavior of $E(k, t)$. In Fig.~\ref{fig7}(a), we plot $E(k,t)$ vs. $k$ for $D=1,2$ at $t=10^3$. As in the CH equation, the step-like variations of $\phi$ at the domain boundaries lead to a Porod tail: $E(k,t) \sim k^{-2}$ in $D=1$, and $E(k,t) \sim k^{-3}$ in $D=2$. The Porod tail is only seen for $k$-values which do not probe the interfacial width $\xi$, i.e., $k \ll \xi^{-1}$. We implement a data-hardening procedure to extend the observation regime of the Porod tail, as done for Fig.~\ref{fig4}.  In this process, we assign $\phi = 1$ for $\phi > 0$, and $\phi = -1$ for $\phi < 0$. Fig.~\ref{fig7}(b) shows the energy spectra corresponding to the hardened data.

\section{Summary and Discussion}
\label{s6}

We now conclude this paper by summarizing our results, and presenting a perspective for future work. We investigated the spectral properties of coarsening models having conserved (CH equation) and non-conserved (TDGL equation) order parameters. Both models incorporate a cubic nonlinearity, which plays a significant role in the evolution of $\phi ({\bf r},t)$ and energy transfers between Fourier modes. We consider initial conditions of the form $\phi({\bf r},0) = \phi_{0} + \delta \phi({\bf r},0)$, where $\delta \phi({\bf r},0)$ denotes random small-amplitude fluctuations. This mimics the disordered system prior to the quench, which occurs at $t=0$. For the CH equation with $\phi_{0} <1/\sqrt{3}$, the system evolves via the emergence and growth of domains. The conservation constraint requires $V^{-1} \int d{\bf r}\phi({\bf r},t) = \phi_{0}$, where $V$ is the volume. In the asymptotic state, we find that $k^2T(k,t) \simeq -2k^2E(k,t)$ for small wavenumbers. The case with $\phi_{0} >1/\sqrt{3}$ can be treated in the linear regime, as long as the system does not form domains. Therefore, this case is analytically tractable and yields the result $k^2T(k,t) \simeq -6\phi^2_0 k^2E(k,t)$ for small non-zero $k$. These results apply for both $D=1$ and $D=2$.

For the TDGL equation with $\phi_{0}=0$, the system again evolves via the emergence and growth of domains. However, there is no conservation constraint in this case. In the asymptotic regime, we find $T(k,t) \simeq -2E(k,t)$ for small $k$, i.e., we obtain the same result as for the CH equation. In the case $\phi_{0}\ne 0$ and $\phi({\bf r},0)$ being uniformly positive or negative, the evolution is substantially different. The average order parameter $a(t) \to \mbox{sgn}(\phi_0)$ as $t \to \infty$. Further, the fluctuations about $a(t)$ decay exponentially with $t$, and can be treated in the linear approximation. The resultant energy transfer equation is $T(k,t) \simeq -6a(t)^2 E(k,t)$ for $k \neq 0$, which is analogous to the result for the linearized CH equation.

In conclusion, spectral energy transfers offer a novel perspective on the coarsening dynamics in the CH and TDGL equations. More generally, these techniques have been used to investigate diverse phenomena which occur on multiple length-scales, e.g., dynamics of oil slicks on the surface of the ocean~\cite{ Binary_berti2005turbulence, james2018turbulence,wang2005two}. These techniques originated in studies of fluid turbulence, but have found diverse application in various scientific and technological problems. The characterization of nonequilibrium systems and pattern formation dynamics remains a major theoretical and experimental challenge. Spectral energy transfers and nonlinear dissipation emerge as valuable diagnostic tools for analyzing these complex systems \cite{verma2021variable,manneville2004turbulence}.

\subsection*{Acknowledgements}

P.K. Yadav is grateful to the UGC, India, for financial support.

\newpage
\bibliography{ref}

\newpage

\end{document}